\documentclass{pasa}%

\title[Origin of magnetic field in PSR J1852+0040]{Origin of magnetar-scale crustal
field in PSR J1852+0040 and ``frozen'' magnetars}
\author[Popov S.B.]{Popov S.B.$^1$\\
\affil{$^1$Sternberg Astronomical Institute, Lomonosov Moscow State
University, Universitetskii pr. 13, 119991 Moscow, Russia
polar@sai.msu.ru}}%
\jid{PASA}
\doi{10.1017/pas.\the\year.xxx}
\jyear{\the\year}

\usepackage[authoryear]{natbib}
\bibpunct{(}{)}{;}{a}{}{,}
\begin{document}%
\begin{abstract}

 We discuss the origin of strong crustal magnetic field in one of
central compact objects (CCOs) ---
a neutron star PSR J1852+0040
in the supernova remnant Kes 79. Taking into account its relatively
long present day spin period
we conclude that the field could not be generated via a dynamo mechanism.
If this neutron star indeed is a magnetar with field submerged during a strong
fall-back episode, then it argues against the dynamo field origin in magnetars.
Otherwise, Kes 79 is not a close relative of normal magnetars.
A discovery of an anti-magnetar with  a  millisecond period and strong crustal field
identifiable, for example, due to large pulse fraction, would be the proof of
the dynamo field origin. Existence of such sources is in correspondence with the
present standard picture of neutron star unification. However, the fraction of 
magnetars with submerged fields can be small --- few percent of the total number
of CCOs.

\end{abstract}
\maketitle%
\section{INTRODUCTION }
\label{sec:intro}

  Magnetars are neutron stars (NSs) whose energy budget is dominated by
dissipation of electric currents supporting magnetic fields. 
Known objects of this type have strong fields: poloidal (in
particular, dipolar)  or/and toroidal.
The origin of these strong fields is still not clear.

 In their foundational paper \cite{dt92} proposed that strong fields are 
enhanced via a dynamo mechanism. This scenario requires rapid initial rotation
of a NS. Field  of the order of $10^{15}$ G can be reached 
if the initial spin is $\sim 1$ msec\footnote{However, as the physical situation
is very complicated, this was never studied in details in realistic computer
simulations.}.
This is a strong prediction that potentially can be tested. 

 \cite{vink} studied supernova remnants (SNRs) with magnetars to check if traces
of additional energy release due to a rapidly spinning down magnetar can be found. 
Indeed, a young magnetar is a very powerful source with a luminosity $L\sim 9.6 \times 10^{48}
\, B_{15}^2 P_{\mathrm{msec}}^{-4}$ erg s$^{-1}$.
Such objects were even considered as central engines of gamma-ray bursts
\citep{usov}. 
Though, no additional energy input in magnetars' SNRs have been found \citep{vink}. 
Still, this result does not exclude the original scenario by \cite{dt92}.
Note also, that alternatives related to fossil fields \citep{ferrario} have
been seriously critisized \citep{spruit}, as just a small fraction of the
total stellar magnetic flux can end in a NS.

 Unfortunately, due to a rapid spin down and field decay
initial periods of magnetars are quickly ``forgotten''.
However, there is a situation, when the initial period can be ``frozen''.
This is related to strong fall-back which results in burying of the magnetic field
\citep{mp95,ho11,vp12,bern12}.
Such a NS --- magnetar inside, anti-magnetar outside, --- can be a Rosetta stone 
for understanding the origin of strong magnetic fields. 
Potentially, a NS in the SNR Kes 79 could be such an object.

 The NS in Kes 79 (PSR J1852+0040) 
belongs to the class of so-called ``anti-magnetars''
\citep{hg10}.
It has a spin period $p=0.105$~s, $\dot p=8.7 \times 10^{-18}$, and so the polar magnetic field
for an orthogonal rotator is about $ 6.2 \times 10^{10}$~G. 
The SNR age is estimated as $\sim 6 \times 10^3$ years.
The source has one peculiar feature: large (64\%) pulse fraction. 
 \cite{sl12} show that large pulse fraction of the NS in Kes 79 can be explained
if its magnetic field in the crust is very strong: few $\times 10^{14}$~G. 
In their study the authors assumed that the strong component of the field is the toroidal one.
Potentially, this opens two possibilities. First, despite its young age the NS in Kes 79 is an
analogue of so-called low-field magnetars (see a review in \citealt{turolla}). 
Second, both components are strong,
and they have been significantly submerged due to a strong fall-back episode
(\citealt{gpz99} called such sources ``hidden magnetars''). 
Below we analyse how this can be used to put constraints on the origin of magnetars' magnetic
field.

\section{ANALYSIS AND ESTIMATES}

The process of fall-back on a newborn magnetar can be a very complicated one.
However, in this section we apply simple formulae to perform 
 a qualitative analysis of the consequences of such a process.

 At first, 
 we recall that accretion on the surface of a NS can be prohibited either by 
a relativistic wind (so-called {\it ejector} stage), or by a rapidly rotating magnetosphere
({\it propeller} stage), see a detailed description of stages, for example, in \cite{lipunov}.
Transition between stages can be formalized in terms of critical spin periods
(back transitions can be not symmetric, and so corresponding critical periods
can be different, also additional conditions can be important, but we do not
discuss here situations  not applicable to the fall-back case).

The first one, $p_\mathrm{E}$, corresponds to the transition
ejector$\rightarrow$propeller.
The second, $p_\mathrm{A}$, --- to the transition propeller$\rightarrow$accretor.
 We
are interested mainly in the  latter one, as the fall-back we are interested
in, is strong enough to switch off the pulsar.


\begin{equation}
p_\mathrm{A}\approx 20 \mu_{32}^{6/7} \left(\frac{M}{M_\odot}\right)^{-5/7} 
\left(\frac{\dot M}{M_\odot \mathrm{yr}^{-1}}\right)^{-3/7} \, \mathrm{msec}.
\end{equation}
Here $\mu$ --- magnetic moment of a NS, $M$ --- its mass, and $\dot M$ ---
accretion rate. We use standard accretion formulae  for the adiabatic
index
$\gamma=5/3$. \cite{che89} is his estimates used $\gamma=4/3$ (a radiation
dominated envelope). However, as can be seen from his eqs. (4.22) and (5.2)
for fields of the magnetar scale the magnetosphere can start to be
important for  $\gamma=4/3$, too. So in the case of a rapidly rotating NS the propeller stage
can appear for reasonable accretion rates also for this value of the
adiabatic index.

 If the initial dipole magnetic field is strong and  the spin period is short, 
then accretion is possible only for a very large accretion rate $\sim 100 M_\odot$~yr~$^{-1}$. 
Such rates are not unexpected. For example, \cite{che89} estimated the
accretion rate for SN1987A as $350 M_\odot$~yr$^{-1}$ (lower values about
tens of solar masses per year also have been reported by other authors). 

\cite{bern12} demonstrated that for such strong fall-back the field 
is rapidly ($\sim 100 $~msec) submerged. 
Then it can diffuse out on a time scale of thousands or tens of thousand years
\citep{ho11,vp12}.
While the field is submerged 
the NS appears as an anti-magnetar. Rotation braking is very slow in this case. 
The spin period is nearly constant, frozen. 
A similar situation appears if the initial dipolar field was small, but the toroidal
 is large. 
In this case
the accretion gates are opened even for lower $\dot M$. 
Discovery of an anti-magnetar 
 with a large crustal field and a millisecond-scale spin period
would be a proof for a dynamo-mechanism field origin. 
Oppositely, PSR J1852+0040 (with its large spin period) 
most probably represents a counter example.

In PSR J1852+0040 the present day spin period is nearly the same as the initial one 
because $\dot p$ 
 is low due to the low field and  the relatively long spin period.
And it is easy to estimate that no spin-down mechanism (magneto-rotational or similar ones,
propeller, interaction with a disc, etc.) 
 can slow the rotation from few milliseconds 
down to 100 msec in a very short time before the field is buried. 

 One of the most effective spin-down mechanisms
 ever discussed was proposed by \cite{shakura}
(see a list of other spin-down formulae for the propeller stage in
\citealt{lp95}):
\begin{equation}
d(I \omega)/dt=\omega \dot M R_\mathrm{A}^2.
\end{equation}
Here $R_\mathrm{A}$ is the magnetospheric (Alfven) radius, 
$\omega=2\pi/p$ --- is the spin frequency, and $I$ --- is
the moment of inertia of a NS.
Eq. (2) can be rewritten as:
\begin{equation}
dp/dt=\dot M R_\mathrm{A}^2 p/I=
\end{equation}
$$ 
=8.6\times 10^{-5} \left(\frac{\dot M}{100 \, M_\odot
\,{\mathrm{yr}}^{-1}}\right)^{3/7} p \, 
\mu_{32}^{8/7} \left(\frac{M}{M_\odot}\right)^{2/7}.
$$
 Despite the exponential spin-down during this stage,
field submergence (as numerical
experiments indicate) happens so rapidly, that the period cannot be
increased by a factor $\sim 100$.  Obviously, normal magneto-dipole spindown
is less effective, and a newborn magnetar cannot be slowed down
significantly due to this mechanism.

We conclude, 
 that a strong crustal toroidal
magnetic field in Kes 79 could not be generated with 
a dynamo mechanism.

\section{DISCUSSION}

\cite{sl12} comment that the field configuration they obtained can be not a unique one
which explains the large pulse fraction of PSR J1852+0040. 
This means that we cannot be sure if the dipolar component of the field 
was strong before the fall-back episode, or not. 

\cite{bern12} 
 mention that the accretion of magnetized matter can modify the magnetic
field of a NS,  reducing the dipolar component while making
the general shape of the field configuration more complicated.
If this is the case for PSR J1852+0040, 
than it can be much different from normal magnetars.
Even if the strength of toroidal field in the crust is high, the total energy of the field can be
orders of magnitude smaller than in magnetars \citep{pons}. Probably, an effective dynamo mechanism
is not necessary in this case. 
Then accretion could start for 100 msec period even for relatively low
accretion rates. 

 High accretion rates normally result in a large total accreted mass,
comparable with the mass of the crust. Roughly, the mass of the crust is
accreted in one hour for the rate $\dot
M=100\, M_\odot\,$~yr$^{-1}$:
$$
\Delta M \sim M_\mathrm{crust} \left(\frac{\dot M}{100\, M_\odot
{\mathrm{yr}}^{-1}}\right) \left( \frac{\Delta t}{1 {\mathrm{hour}}}\right). 
$$
 The total accreted mass can be up to few tenths of solar mass
\citep{che89}.
 A NS which experienced such strong fall-back episode can so much more
massive that it follows a different cooling history.
In all scenarios massive NSs are colder \citep{yakovlev,page,blaschke}.
As far as the NS in Kes 79 is a hot source, then an additional heating mechanism is required.
It can be easily provided due to rapid magnetic field decay in the crust, as
the field is confined in a narrow layer (see, for
example, \citealt{gpz99}). 

 It would be important to estimate for PSR J1852+0040 the total amount of accreted matter. 
 This can allow the calculation of
the maximum accrection rate during the fall-back.
Then all estimates above can be made in a more robust way.

As a determination of initial spin periods for known magnetars\footnote{See
the list in the on-line catalogue at the McGill university
http://www.physics.mcgill.ca/~pulsar/magnetar/main.html.} is impossible, a
promising way to get solid arguments in favour of the dynamo mechanism
hypothesis is to find a real ``frozen magnetar''. In the modern picture of
unified NS population (see, for example, \citealt{geppert} for description
of different possibilities and evolutionary histories, especially their Fig.
2 for evolutionary tracks) such sources are quite natural. However, they
must be relatively rare. 

 Magnetars form up to 10\% of all NSs as both observations and population
synthesis studies indicate (see \citealt{gillheyl2007}, and 
also  \citealt{popov} for a unified desription of magnetars, pulsars and
cooling NSs, and references therein).  
If  we assume that the submergence of the magnetic field happens
in magnetars as often, as in other
NSs, and that CCOs are NSs with submerged fields,
then the number of ``frozen'' magnetars is $\lesssim 10$\% of the number of CCOs. 
 However, it is natural to assume that
the larger the fields, the more difficult it is to bury them.
Then, the fraction of ``frozen magnetars'' is even lower. 
In addition, if magnetars are born in binary systems where spin of the
progenitor was increased due to interaction with the companion
\citep{pp2006,bp2009}, then the amount of fall-back 
 can be smaller as a significant part of the outer layers
of the star are already removed (this
was mentioned, for example, by \citealt{che89}).
 The problem of identification (the detectability of millisecond periods for
complicated field topology in the crust, pulse profiles, etc.)
also can prevent easy discovery of ``frozen magnetars'' (but
see a recent discussion in \citealt{perna13}). 
 On the other hand, an additional energy source -- field decay -- can help
such sources to stay bright for a longer time, which favours their
discovery \citep{heyl1998}.

\section{CONCLUSIONS}

 The CCO in Kes 79 has a relatively long spin period (0.1 s) and might have a
strong crustal field \citep{sl12}.
We argue that this field could not be generated by a dynamo mechanism
as, on one hand, known mechanisms of NS spin down applicable in the case of this source
cannot slow down rotation form $\sim$~msec to the present day period before the
field is buried, on the other --- the present day low field  excludes also a
significant spin-down during the lifetime of the source. 
Discovery of a ``frozen magnetar'' --- a CCO-like source
with a millisecond period and 
 indications of a strong crustal field, -- would be
a strong argument in favour of the dynamo mechanism scenario by
\cite{dt92}.

\begin{acknowledgements}
I thank participants of the Madrid-2013 conference ``Fast and furious'',
in particular dr. Daniele Vigan\'o, for discussions.
Special thanks to prof. D. Page, who made usuful comments on the  draft of
the manuscript. Comments by an unknown referee helped to improve the
paper. 
This work was supported by the RFBR grant 12-02-00186.
\end{acknowledgements}


\bibliographystyle{apj}
\bibliography{popovkes79}

\end{document}